\documentclass[aps,prl,twocolumn,floatfix,showpacs]{revtex4-1}
\usepackage{epsfig,subfigure,amssymb,amscd,amsmath,graphicx,amsfonts}
\usepackage{times,longtable,braket,dsfont,bbold,txfonts}
\usepackage[usenames]{color}

\definecolor{pinegreen}{rgb}{0.0, 0.47, 0.44}

\begin{document}
 
\title{Density of states at disorder-induced phase transitions in a multichannel Majorana wire}

\date{\today}

\author{Maria-Theresa Rieder, Piet W. Brouwer}

\affiliation{Dahlem Center for Complex Quantum Systems and Fachbereich Physik, Freie Universit\"{a}t Berlin, Arnimallee 14, 14195 Berlin, Germany}

\begin{abstract}
	An $N$-channel spinless p-wave superconducting wire is known to go through a series of $N$ topological phase transitions upon increasing the disorder strength. Here, we show that at each of those transitions the density of states shows a Dyson singularity $\nu(\varepsilon) \propto \varepsilon^{-1}|\ln\varepsilon|^{-3} $, whereas $\nu(\varepsilon) \propto \varepsilon^{|\alpha|-1}$ has a power-law singularity for small energies $\varepsilon$ away from the critical points. Using the concept of ``superuniversality'' [Gruzberg, Read, and Vishveshwara, Phys.\ Rev.\ B {\bf 71}, 245124 (2005)], we are able to relate the exponent $\alpha$ to the wire's transport properties at zero energy and, hence, to the mean free path $l$ and the superconducting coherence length $\xi$.
\end{abstract}


\maketitle

\textit{Introduction.} Though stable against moderate amounts of disorder, topological phases are typically susceptible to strong disorder. This is particularly true for topological phases in one and two dimensions, for which strong disorder eventually leads to a localization of all electronic states. However, there are examples in which the effect of disorder may not be simply the transition from the topological into a topologically trivial localized phase, but more diverse physics appears.

The most prominent such example is the quantum Hall effect where disorder is an essential element needed to stabilize the topological phase and, hence, to explain the quantization of the conductance \cite{Chalker1999}.
In the context of time-reversal invariant topological insulators, a topologically trivial system may be driven into a nontrivial phase by disorder, as it happens for topological Anderson insulators \cite{Li2009,Groth2009,Guo2010a,Guo2010b}. Also, when disorder preserves certain symmetries on the average, the disorder itself may drive a topological insulator into a new type of topological phase, the so-called ``statistical topological insulator'' \cite{Fu2012,Fulga2014}. Topological superconductors, finally, can display thermal metal \cite{Fulga2012} or glassy phases \cite{Crepin2014} or enter a topologically nontrivial phase upon increasing disorder strength \cite{Adagideli2014}.

An example where disorder leads to a particularly rich phase diagram is that of a multichannel spinless superconducting wire. In Ref.\ \onlinecite{Rieder2013}, the authors, together with Adagideli, showed that upon increasing the disorder strength such a wire goes through a series of topological phase transitions, alternating between states with and without a Majorana bound state at the wire's end \cite{Rieder2013}. For a wire with $N$ transverse channels, there are $N$ such transitions, which take place at mean-free path
\begin{equation}
  l^{(n)}_{\rm crit} = 
  \frac{n \xi}{N+1},\ \ n=1,2,\ldots,N,
  \label{eq:lcrit}
\end{equation}
where $\xi$ the superconducting coherence length \cite{foot1}.

Whereas Ref.\ \onlinecite{Rieder2013} identified the location of the topological phase transitions, it did not discuss the system's spectral and transport properties in the vicinity of the critical point. A theoretical framework in which this question can be addressed was provided by Gruzberg, Read, and Vishveshwara \cite{Gruzberg2005}, who argued that there exists a ``superuniversality'', according to which all disorder-induced critical points in (quasi-)one dimension are of the same type as the critical point in the one-dimensional non-superconducting chiral class. For the chiral class, at the critical point the density of states displays a Dyson singularity \cite{Dyson1953} $\nu(\varepsilon) \propto \varepsilon^{-1}|\ln\varepsilon|^{-3}$, whereas away from the transition, a power law $\nu(\varepsilon)\propto \varepsilon^{|\alpha|-1}$ is expected as $\varepsilon \to 0$, where $\alpha$ is a dimensionless parameter that measures the distance to the critical point. Wavefunctions and transmission probabilities (in the case of a system coupled to source and drain leads) also have universal statistics, parameterized by the same parameter $\alpha$. The density-of-states singularity and the associated wavefunction or transmission statistics occur in a wide range of physical systems, including lattice models with random hopping \cite{Theodorou1976,Eggarter1978}, quantum $XY$ chains \cite{Fisher1994}, narrow-gap semiconductors \cite{Ovchinnikov1977}, dimerized polymer chains \cite{Su1980,Jackiw1983,Rice1982}, and single-channel spinless superconductors \cite{Motrunich2001,Brouwer2011b,Gruzberg2005}. Following the reasoning of Ref.\ \onlinecite{Gruzberg2005} the same critical behavior is expected to apply to the multichannel Majorana wire. It remains to express the dimensionless parameter $\alpha$ in terms of the model parameters, the mean free path $l$ and the coherence length $\xi$. 

\textit{Multichannel Majorana wire.}	We consider a disordered spinless p-wave superconducting wire in two dimensions, in a wire geometry with width $W$ and length $L \to \infty$. The Hamiltonian for such a system has the form 
\begin{align}
  H= \left[ \frac{p^2}{2m} + V(x,y) - \mu \right] \sigma_{z}+
  \Delta'_x p_{x} \sigma_{x}+\Delta'_y p_{y} \sigma_{y},
  \label{eq:HD}
\end{align}
where $0 < x < L$ and $0 < y < W$ are longitudinal and transverse coordinates, respectively, the matrices $\sigma_{x,y,z}$ are Pauli matrices in electron/hole space, $\mu$ is the chemical potential, $m$ the electron mass, $\Delta'_{x,y}$ are the $p$-wave superconducting pairing terms in the longitudinal and transversal directions, and $V(x,y)$ is the disorder potential, which is characterized through the elastic mean free path $l$. The number of channels $N$ is defined as the number of propagating modes at the Fermi level in the absence of superconductivity. The model (\ref{eq:HD}) is an effective low-energy description of a system in which the superconducting correlations come from proximity coupling to a nearby $s$-wave spinfull superconductor \cite{Fu2008,Lutchyn2010,Oreg2010,Cook2011,Duckheim2011,Rieder2012}, so that no self-consistency condition for $\Delta'_x$ and $\Delta'_y$ needs to be accounted for. The Hamiltonian $H$ of Eq.\ (\ref{eq:HD}) has no other symmetries than particle-hole symmetry, implying that the system is in symmetry class ``D'' according the Cartan classification \cite{Altland1997,Ryu2010,Evers2008}.

For thin wires $W \ll \xi$, with the superconducting coherence length $\xi = \hbar/m \Delta'_x$, the term $\Delta'_y p_{y} \sigma_{y}$ only has a small effect on the wavefunctions and the spectrum and can be treated in perturbation theory. Without it, $H$ obeys the chiral symmetry $\sigma_y H \sigma_y = - H$ \cite{Tewari2012,Kells2012}. In the Cartan classification this corresponds to symmetry class ``BDI''. Since the presence of the chiral symmetry significantly simplifies the calculation of the Majorana end states, Ref.\ \onlinecite{Rieder2013} first analyzes the model (\ref{eq:HD}) without the term $\Delta'_y p_y \sigma_{y}$. Here we take the same approach.

In the absence of disorder, and without the term $\Delta'_y p_y \sigma_{y}$, there are $N$ Majorana bound states at each end of the wire, with a wavefunction that decays exponentially with decay length $\xi$ upon moving away from the wire's end. With disorder, but still in symmetry class ``BDI'', a suitable basis of transverse channels can be chosen, such that the wavefunction envelope of the $n$th Majorana state at the wire's left end decays as \cite{Rieder2013}
\begin{equation}
  \psi^{(n)}_{L/R}(x) \propto e^{- x/\xi + xn/(N+1)l},\ \
  n=1,2,\ldots,N,
\end{equation}
where $l$ is the mean free path for scattering from the disorder potential $V$. At the critical disorder strengths $l^{(n)}_{\rm crit}$ the wavefunction of the $n$th Majorana end state becomes delocalized, indicating a (topological) phase transition. At the phase transition, the $n$th Majorana end states at the two ends of the wire hybridize and annihilate. Increasing the disorder strength therefore leads to a series of $N$ topological phase transitions in which the $N$ Majorana bound states at the wire's end disappear one by one until the system reaches the topologically trivial state without Majorana end states. 

The effect of including the term $\Delta'_y p_y \sigma_y$ is that Majorana end states at the same end of the wire can annihilate pairwise. Hence, one Majorana end state remains if the number of Majorana end states before including $\Delta'_y p_y \sigma_y$ was odd, and no Majorana end state remains if the number of Majorana end states was even. Thus, for the full Hamiltonian (\ref{eq:HD}), the number of Majorana end states alternates between zero and one upon increasing the disorder strength, with the transitions approximately (with corrections that vanish in the limit $W/\xi \to 0$) taking place at the critical disorder strengths specified in Eq.\ (\ref{eq:lcrit}).

In Ref.\ \onlinecite{Rieder2013} this conclusion was reached by attaching source and drain leads to the Majorana wire with Hamiltonian (\ref{eq:HD}) and formally mapping the scattering matrix of this problem to that of the disordered wire in the normal state (at a slightly renormalized chemical potential). In this mapping, the total quasiparticle conductance $T$ of the Majorana wire in the limit $L \gg \xi$, $N l$ can be easily expressed in terms of the transmission eigenvalues $\tau_n$ of the disordered wire in the normal state
\begin{equation}
  T = \sum_{n=1}^{N} \frac{(\tau_n/2) e^{2 L/\xi}}{[1 + (\tau_n/4) e^{2 L/\xi}]^2}.
\end{equation}
The probability distribution of the transmission eigenvalues $\tau_n$ for large $L$ and weak disorder is known in the literature \cite{Beenakker1997},
\begin{equation}
  \langle \log \tau_n \rangle = \frac{2 n L}{(N+1) l},\ \
  \mbox{var}\, \log \tau_n = \frac{4 L}{(N+1)l}.
  \label{eq:taudistr}
\end{equation}
For a mean free path $l$ near the critical value $l_{\rm crit}^{(n)} = n \xi/(N+1)$, the quasiparticle transmission is dominated by the $n$th transmission eigenvalue $\tau_n$. Using the parameterization 
\begin{equation}
  T = 1/\cosh^2 z, \label{eq:Tz}
\end{equation}
one finds
\begin{equation}
  \langle z \rangle = \left( \frac{l^{(n)}_{\rm crit}}{\xi} - \frac{l}{\xi} \right) \frac{L}{l},\ \
  \mbox{var}\, z = \frac{L}{(N+1)l}.
  \label{eq:zstat}
\end{equation}
Indeed, at the critical disorder strength (and at the critical disorder strength only) quasiparticle wavefunctions are delocalized throughout the sample \cite{Akhmerov2011}.

The mapping between the scattering matrices of the disordered wire with and without superconductivity that was used in Ref.\ \onlinecite{Rieder2013} exists for zero energy only. For that reason, Ref.\ \onlinecite{Rieder2013} could not access the density of states $\nu(\varepsilon)$ of the multichannel Majorana wire in the vicinity of the critical points. We now show how the density of states can be obtained from the transmission statistics of Eq.\ (\ref{eq:Tz}) and (\ref{eq:zstat}) using the ``superuniversality'' argument of Ref.\ \onlinecite{Gruzberg2005}.

\textit{Mapping to one-dimensional model with chiral symmetry}. According to the ``superuniversality'' argument of Gruzberg, Read, and Vishveshwara \cite{Gruzberg2005}, the quasiparticle transmission distribution $T$ and the density of states $\nu(\varepsilon)$ in the vicinity of the critical point should be the same as that of a one-dimensional disordered wire in the chiral symmetry class. (In this respect, the three chiral classes BDI, AIII, and CII are interchangeable.) Such systems have been analyzed abundantly in the literature, see, {\em e.g.}, Refs.\ \onlinecite{Comtet1995,Brouwer1998,Brouwer2001,Theodorou1976,Eggarter1978,Stone1981,Titov2001}, and we here summarize the main results of relevance to the present problem.

A prototype of the disordered wire with chiral symmetry in one dimension is described by the Hamiltonian \cite{Comtet1995}
\begin{equation}
  H_{\rm chiral} = -v_{\rm F} p \sigma_z + w(x) \sigma_x,
  \label{eq:Hchiral}
\end{equation}
where $v_{\rm F}$ is the Fermi velocity and $w$ is a random potential with mean $\langle w(x) \rangle = (\hbar v_{\rm F} \alpha)/(2 \bar{l})$ and variance $\langle w(x) w(x') \rangle = (\hbar^2 v_{\rm F}^2/\bar{l}) \delta(x-x')$. The parameter $\alpha$ measures the distance to the critical point; $\bar l$ is the mean free path in this system.
In the vicinity of the critical point, the transmission $T = 1/\cosh^2 z$ of such a disordered one-dimensional wire of length $L$, coupled to ideal source and drain leads has a distribution given by
\begin{equation}
  \langle z \rangle = \alpha \frac{L}{2\bar{l}}, \ \ 
  \mbox{var}\, z = \frac{L}{\bar{l}}.
  \label{eq:zstat2}
\end{equation}
The density of states $\nu(\varepsilon)$ has a singularity at zero energy, which is best described through the integrated density of states
\begin{equation}
  N(\varepsilon) = \int_0^{\varepsilon} \nu(\varepsilon) \, .
\end{equation}
Extending the zero-$\alpha$ calculation of Ref.\ \onlinecite{Titov2001} to the case of nonzero $\alpha$, we find the integrated density of states
(see Supplementary Material for details)
\begin{equation}
  N(\varepsilon) = \frac{L}{2\bar{l} |K_{\alpha/2}(2 i \varepsilon v_{\rm F}/\hbar \bar l)|^2} \,,
  \label{eq:NK}
\end{equation}
where $K_{\nu}(x)$ is the Bessel function of the second kind.
For $\alpha = 0$ Eq.\ (\ref{eq:NK}) reproduces the Dyson singularity $\nu(\varepsilon) \propto 1/[\varepsilon \ln^3 (\varepsilon v_{\rm F}/\hbar \bar l)]$, whereas for nonzero $\alpha$ one has the asymptotic dependence $\nu(\varepsilon) \propto |\varepsilon|^{|\alpha| - 1}$. Near the critical point $\alpha = 0$ Eq.\ (\ref{eq:NK}) is to be preferred over the asymptotic power-law dependence, because it applies to a much wider range of energies than the simple asymptotic power law $\nu(\varepsilon) \propto |\varepsilon|^{|\alpha| - 1}$.

Comparing Eqs.\ (\ref{eq:zstat}) and (\ref{eq:zstat2}), one immediately identifies
\begin{align}
  \alpha_n &= \frac{2 (N+1)}{\xi} \left( l_{\rm crit}^{(n)} - l \right),\ \
  \bar l = (N+1) l,
  \label{eq:alpharesult}
\end{align}
as the dimensionless distance to the $n$th critical point for the disordered multichannel Majorana wire, and the equivalent mean free path in the model (\ref{eq:Hchiral}), respectively. The density of states and transmission statistics are governed by the distance to the closest critical point,
\begin{equation}
  |\alpha| = \min_{n=1}^{N} |\alpha_n|.
  \label{eq:dmin}
\end{equation}

\textit{Numerics}. We now compare our predictions to numerical simulations of a disordered multichannel Majorana wire. For technical reasons, we first present numerical calculations for a slight variation of the model (\ref{eq:HD}), in which the Majorana wire is represented by $N$ coupled one-dimensional channels with Hamiltonian
\begin{align}
  \label{eq:equiv_Ham}
  H_{mn} = \delta_{mn}  
  \left[ 
  \left(- {\hbar^2 \over 2m} \partial_x^2 -\mu  \right) \sigma_z
  -i \Delta'_x  \partial_x \sigma_x
  \right]
  + u_{mn}(x) \sigma_z,
\end{align}
with a disorder term $u_{mn}(x)$ that has zero mean and variance
\begin{equation}
  \braket{u_{ij}(x)u_{kl}(x')}={(\hbar v_{\rm F})^2 \over l (N+1)} \delta(x-x') \left(\delta_{ik} \delta_{jl} + \delta_{il} \delta_{jk}\right),
  \label{eq:ustat}
\end{equation}
$l$ being the mean free path. The technical advantage of Eq.\ (\ref{eq:equiv_Ham}) is that the normal-state distribution (\ref{eq:taudistr}) of the transmission eigenvalues also holds up to moderately strong disorder strengths, so that numerical calculations can be performed for (comparatively) smaller system sizes. The Hamiltonian (\ref{eq:equiv_Ham}) anticommutes with $\sigma_y$, {\em i.e.}, it is in symmetry class BDI.

\begin{figure}[t]
	\centering
		\includegraphics[width=.45\textwidth]{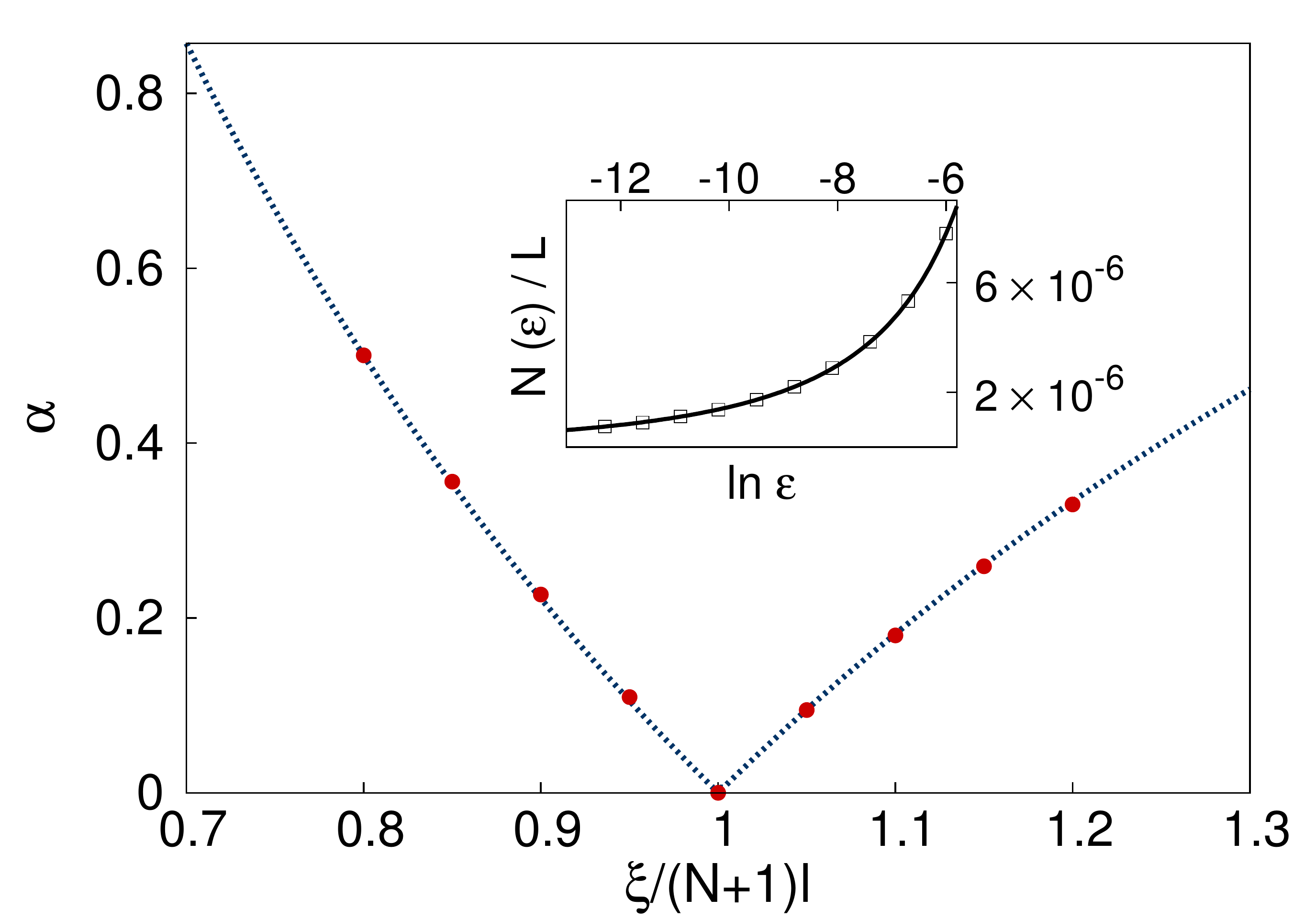}
	\hspace{5mm}	
		\includegraphics[width=.45\textwidth]{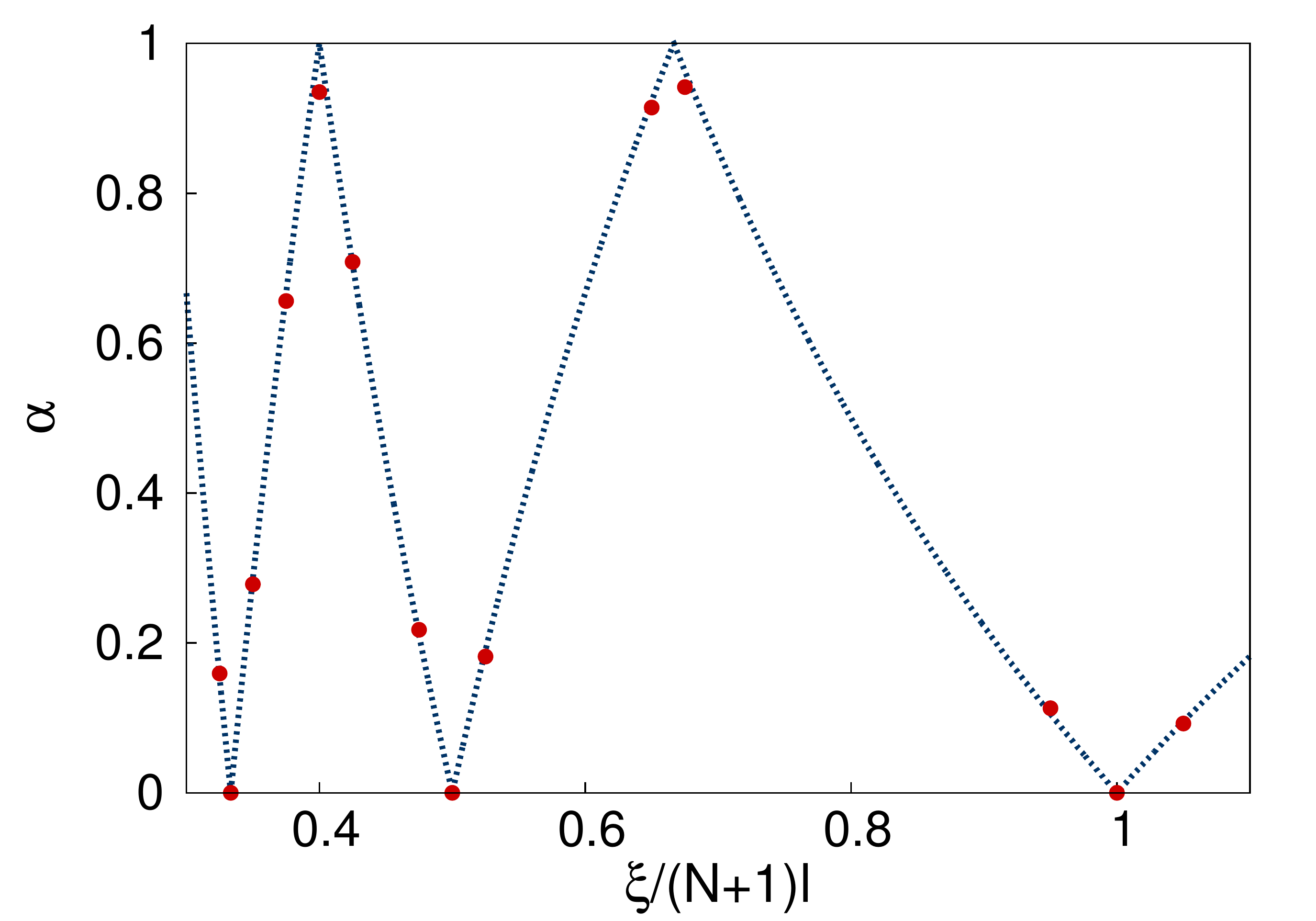}		\caption{(Color online) Dimensionless distance to the critical point as a function of the ratio $\xi/l$ of superconducting coherence length and mean free path. The red dots show the exponent $\alpha$ obtained by fitting the numerically computed integrated density of states to the functional form in  Equation~\eqref{eq:NK} for a single-channel wire (top) and a three-channel wire (bottom). The dashed curve is the exponent expected from the mapping onto a single channel hopping model, see Eq.~\eqref{eq:dmin}. Inset: Example of a fit of the integrated density of states normalized by the wire length as a function of energy. The squares show the numerically obtained data and the continuous curve is the analytical result, Eq.~\eqref{eq:NK}, using the value $\alpha = 0.0906$ obtained from the fitting procedure. The value of $\bar l$ can be obtained directly from the model parameters and need not be fitted, see Eqs.\ (\ref{eq:alpharesult}) and (\ref{eq:ustat}).}
	\label{fig:single_channel}
\end{figure}

In order to determine the density of states, we couple one end of the $N$-channel wire to an ideal lead, keeping the other end closed. Following the method of Ref.\ \onlinecite{Brouwer2011a} we calculate the wire's scattering matrix $S(\varepsilon,L)$ as a function of the length $L$ of the disordered wire. The integrated density of states $N(\varepsilon)$ can be obtained by numerically integrating the relation 
\begin{equation}
  \frac{\partial N(\varepsilon)}{\partial L} =
  \frac{1}{2 \pi} \mbox{Im}\, \frac{\partial \log \det S(\varepsilon)}{\partial L}.
\end{equation}
The integrated density of states obtained this way can be fitted to the functional form (\ref{eq:NK}), which allows us to obtain the dimensionless parameter $\alpha$ as a function of the disorder strength. Results for Majorana wires with $N=1$ and $N=3$ are shown in Fig.\ \ref{fig:single_channel}. The agreement is excellent and holds throughout the entire range of disorder strengths, including points far away from the critical disorder strengths.

\begin{figure}[t!]
	\centering
		\includegraphics[width=.45\textwidth]{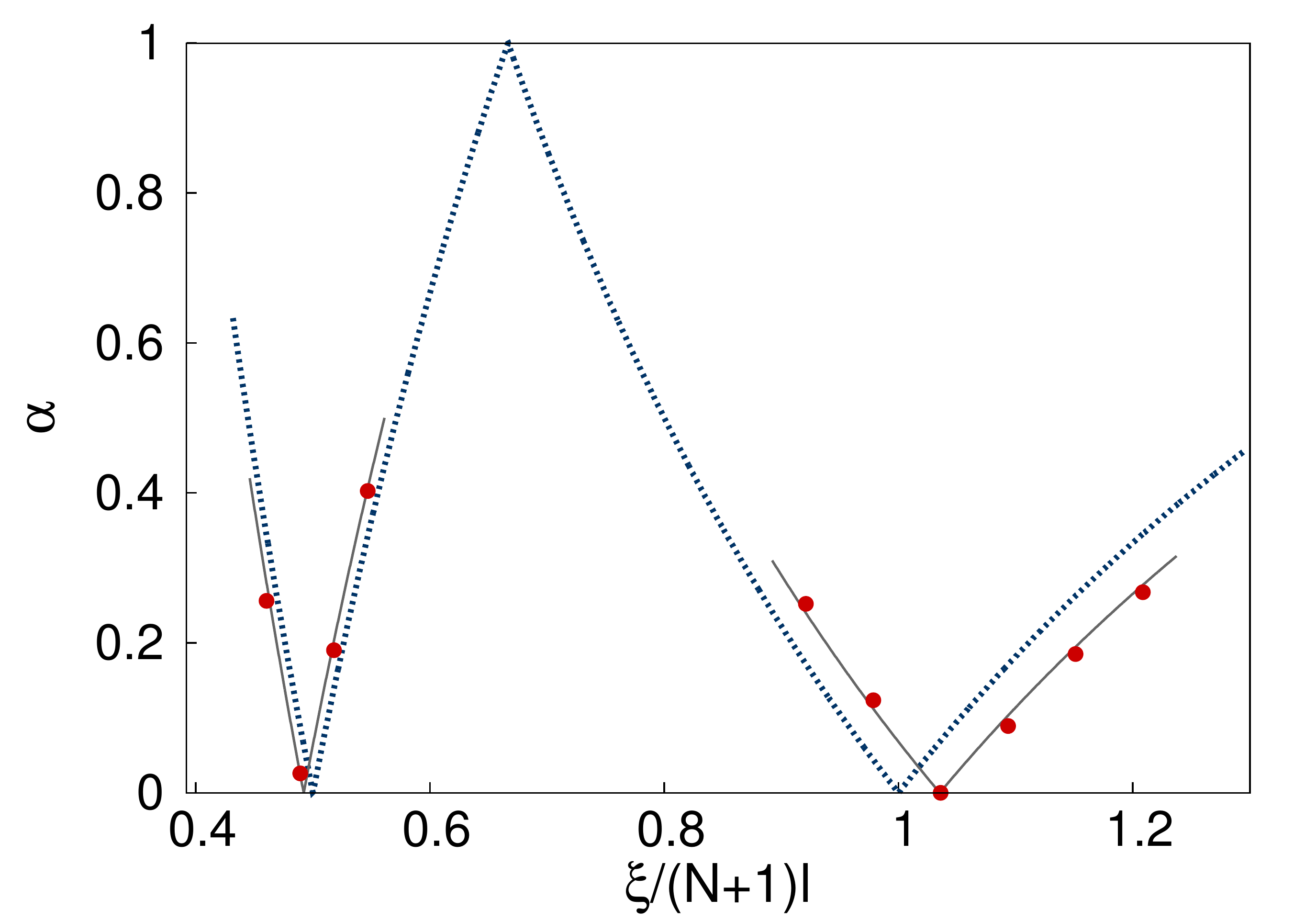}
	\caption{(Color online) Dimensionless distance $\alpha$ to the critical point for the model (\ref{eq:HD}) with $\Delta_x' = \Delta_y'$ and $N=2$ transverse channels. The dashed line is the analytical prediction (\ref{eq:dmin}). The continuous line is the analytical prediction corrected for the slight shift of the critical disorder strength at the second phase transition.
	\label{fig:D_2channel}}
\end{figure}

We have also performed numerical calculations for the two-dimensional Hamiltonian (\ref{eq:HD}) in a strip geometry. We choose the two pairing terms $\Delta'_x$ and $\Delta'_y$ to be equal. Since such a system is no longer in class BDI, we expect slight deviations in the quantitative estimates of the critical disorder strength and the dimensionless distance to the critical point. The numerical results for a wire with $N=2$ indeed show a slight deviation of the critical disorder strength at the second phase transition, although, within the accuracy of our numerical calculations, no deviation for the dimensionless distance $\alpha$ can be discerned, see Fig~\ref{fig:D_2channel}.

\textit{Conclusion}. We have investigated the density of states of a multichannel spinless superconducting wire, as it goes through a series of disorder-driven topological phase transitions. Using the concept of ``superuniversality'' of Gruzberg, Read, and Vishveshwara \cite{Gruzberg2005}, we could establish a relation between the known quasiparticle transmission statistics at zero energy and the singular contribution to the density of states at finite energies. A comparison with a numerical solution of the problem is in excellent agreement with these analytical results. Our results are a powerful demonstration of the concept of superuniversality, showing that in one dimension, as well as in quasi one dimension, the scaling relations for the density of states remain valid across boundaries between symmetry classes.

We gratefully acknowledge discussions with Christopher Mudry and Alexander Altland. This work is supported by the Alexander von Humboldt Foundation in the framework of the Alexander von Humboldt Professorship, endowed by the Federal Ministry of Education and Research.

{\em Appendix.}
To the best of our knowledge, Eq.\ (\ref{eq:NK}) is not known in the literature, although it can be derived rather quickly by adapting existing calculations of the density of states in a wire with chiral symmetry at the critical point $\alpha=0$. Here we take Ref.\ \onlinecite{Titov2001} as our starting point, where the density of states was calculated from the stationary distribution $P(x)$ of the reflection eigenvalue $R = \tanh^2(x)$ of a wire with Hamiltonian (\ref{eq:Hchiral}), evaluated at the {\em imaginary} energy $\varepsilon = -i \omega$, $\omega > 0$, and in the limit of a large wire length $L$. In Ref.\ \onlinecite{Titov2001} this distribution is found as the stationary solution of the Fokker-Planck equation
\begin{equation}
  \label{eq:FP_absorption}
  \frac{\partial P(x)}{\partial L} = 
  \frac{\partial}{\partial x}
  \left[ \frac{\omega}{v_{\rm F}} \sinh 2 x +
  \frac{1}{2 \bar l} J \frac{\partial}{\partial x} J^{-1} \right] P(x),
\end{equation}
where $J$ is a Jacobian which, for the case of a one-dimensional wire with chiral symmetry takes the value $J=1$ at the critical point $\alpha=0$. Solving Eq.\ (\ref{eq:FP_absorption}) gives the stationary solution
\begin{equation}
  \label{eq:jpd_absorption}
  P(x) = {1 \over Z(a)} |J| e^{-a \cosh 2 x},
\end{equation}
with $a= \omega l/\hbar v_{\rm F}$ and $Z(a)$ a normalization factor. The key result of Ref.\ \onlinecite{Titov2001} is a general relation between the integrated density of states $N(\varepsilon)$ and this normalization factor,
\begin{equation}
  N(\varepsilon) = \frac{L}{\pi \bar l}\, \mbox{Im}\,
  \left[ a \frac{\partial}{\partial a} \ln Z(a) \right]_{a \to -i \bar l \varepsilon/\hbar v_{\rm F}}.
\end{equation}

The calculation of Ref.\ \onlinecite{Titov2001} is easily generalized to the case $\alpha \neq 0$: Nonzero $\alpha$ gives rise to a constant drift term in the Fokker-Planck equation (\ref{eq:FP_absorption}) \cite{Brouwer1998} or, equivalently, an exponential factor in the Jacobian $J$,
\begin{equation}
  J = e^{-\alpha x}.
\end{equation}
The stationary solution and the integrated density of states are then obtained in the same way as described above. One finds
\begin{equation}
  Z(a) = K_{\alpha/2}(a),
\end{equation}
from which the result (\ref{eq:NK}) follows directly.

\appendix

\end{document}